\documentclass[12pt,oneside, a4paper]{article}
\pdfoutput=1

\ifx\pdfoutput\undefined
\usepackage[dvips,bookmarks=false]{hyperref}	
\else
\usepackage{hyperref}	
\fi
\hypersetup{colorlinks,bookmarksopen,bookmarksnumbered,citecolor=blue,
linkcolor=black,pdfstartview=FitH,urlcolor=blue}

\usepackage{graphicx}
\usepackage{amssymb}
\usepackage{cite}
\usepackage{bm}
\usepackage{indentfirst}
\usepackage{amsmath}
\usepackage{hhline}
\usepackage{multirow}

\begin{document}

\begin{titlepage}

\begin{flushright}
KANAZAWA-14-06\\
IPPP-14-39\\
DCPT-14-78
\end{flushright}

\begin{center}

\vspace{1cm}
{\large\bf 
 Impact of Semi-annihilation of $\mathbb{Z}_3$ Symmetric Dark Matter\\
 with Radiative Neutrino Masses}
\vspace{1cm}

\renewcommand{\thefootnote}{\fnsymbol{footnote}}
 Mayumi Aoki$^1$\footnote[1]{mayumi@hep.s.kanazawa-u.ac.jp} 
and 
Takashi Toma$^2$\footnote[2]{takashi.toma@durham.ac.uk}
\vspace{5mm}

{\it%
$^1${Institute for Theoretical Physics, Kanazawa University, Kanazawa
 920-1192, Japan}
$^2${Institute for Particle Physics Phenomenology\\ 
University of Durham,  Durham DH1 3LE, United Kingdom}}

\vspace{8mm}

\abstract{
 We investigate a $\mathbb{Z}_3$ symmetric model with two-loop radiative neutrino
 masses. Dark matter in the model is 
 either a Dirac fermion or a complex scalar as a result of an unbroken
 $\mathbb{Z}_3$ symmetry. In addition to standard annihilation processes,
 semi-annihilation of the dark matter contributes to the relic density.
 We study the effect of the semi-annihilation in the model and find that 
 those contributions are important to obtain the observed relic density. 
 The experimental signatures in dark matter searches are also discussed,
 where some of them are expected to be different from the signatures of
 dark matter in $\mathbb{Z}_2$ symmetric models.
 }

\end{center}
\end{titlepage}

\renewcommand{\thefootnote}{\arabic{footnote}}
\setcounter{footnote}{0}

\setcounter{page}{2}

\section{Introduction}
Non-zero small masses and mixings of the neutrinos have been confirmed by neutrino oscillation
experiments such as solar, atmospheric, accelerator, and reactor neutrino
experiments. All the data have been gathered 
together and the global fit of neutrino parameters has been done in
ref.~\cite{GonzalezGarcia:2012sz}. 
However the generation mechanism of neutrino masses is not known yet. 
It could be the canonical seesaw mechanism~\cite{Minkowski:1977sc,
Yanagida:1979as, GellMann:1980vs} in which the masses 
are derived with superheavy right-handed neutrinos, but 
verifying this experimentally is difficult. 

On the other hand, the evidence for Dark Matter (DM) has been inferred from many observations, 
such as the rotation curves of spiral
galaxies~\cite{Begeman:1991iy}, the Cosmic Microwave
Background~\cite{Ade:2013zuv} and the collision of the bullet 
cluster~\cite{Clowe:2006eq}. However, properties of the DM particle
like its mass and interactions are not known. One of the most promising DM candidates is
a Weakly Interacting Massive Particle (WIMP), which is thermally produced by
decoupling from the thermal bath in the early universe. To reveal the
nature of DM, various experiments including direct, indirect and
collider searches are being operated. 

Radiative seesaw models are one interesting possibility that address both of the issues above. 
Owing to the loop suppression, in these models the small neutrino masses are 
naturally obtained from TeV scale physics. 
 In such models, the neutrino phenomenology often correlates
with DM physics since a discrete symmetry like $\mathbb{Z}_2$ parity forbids
tree-level neutrino mass terms and also stabilizes the DM candidate. 
In addition to the representatives of well-known radiative models~\cite{Ma:2006km,
Zee:1985id, Babu:1988ki, Krauss:2002px, Aoki:2008av}, 
many different models have been proposed and analyzed, see for example~\cite{Aoki:2010ib, Kanemura:2011mw, Aoki:2013gzs, Kajiyama:2013rla,
Kanemura:2013qva, Baek:2013fsa, Baek:2014awa, Ahriche:2014cda,
Ahriche:2014oda,Hehn:2012kz, Hirsch:2013ola, Restrepo:2013aga,
Kanemura:2014rpa, Okada:2013iba, Kajiyama:2013zla, Lindner:2013awa,
Kanemura:2011vm, Okada:2014nsa, Gustafsson:2012vj, Law:2013saa,
MarchRussell:2009aq, Dev:2012sg, Dev:2012bd}. 
Although the $\mathbb{Z}_2$ symmetry is introduced to
stabilize the DM candidate in most of the models, other symmetries are
possible. The second simplest symmetry is $\mathbb{Z}_3$ 
and the properties have been studied~\cite{D'Eramo:2010ep,
Belanger:2012vp, D'Eramo:2012rr, Belanger:2012zr, Ko:2014nha,
Belanger:2014bga}. In these models, semi-annihilation of DM such as 
$\chi\chi\to\chi^\dag X$ plays an important role in evaluating the DM
relic density, where $\chi$ is $\mathbb{Z}_3$ charged DM and $X$ is a
Standard Model (SM) particle.\footnote{Semi-annihilation of DM in models without a
$\mathbb{Z}_3$ symmetry have been discussed in the framework of vector
boson DM~\cite{Hambye:2008bq, Arina:2009uq, Khoze:2014xha, Boehm:2014bia}, 
and in multi-component DM scenarios~\cite{Ivanov:2012hc,
Aoki:2012ub, Belanger:2014bga}.}

In this paper, we consider a simple $\mathbb{Z}_3$ symmetric model with
the radiative neutrino masses at the two-loop level, which has been
proposed in ref.~\cite{Ma:2007gq}. The DM candidate in the model is
either a Dirac fermion or a complex scalar. 
However, in ref.~\cite{Ma:2007gq}, it has been mentioned that both of them are unsuitable
as DM candidate 
owing to 
the inconsistency with the experimental data of
direct detection 
or Lepton Flavor Violation (LFV).
Nevertheless, we will show that this conclusion changes when semi-annihilation processes
are taken into account,
which are specific properties of the
$\mathbb{Z}_3$ symmetric model.
In the next section, the model is introduced and the two-loop induced neutrino
masses are evaluated. Some experimental
constraints are also discussed.
In Section~\ref{sec:3}, the DM properties are investigated in detail for
the Dirac fermion DM and the complex scalar DM, respectively.
We discuss two cases for new Yukawa coupling with charged leptons $y^\nu$;
small $y^\nu$ without any specific flavor structure and large $y^\nu$
with a specific flavor structure. 
In the calculation of the DM relic density, 
the semi-annihilation processes will give considerable effects. In
particular for the Dirac DM, the severe constraints 
from LFV and the relic density of DM can be
consistent due to the effects of semi-annihilations. 
Detectability of DM is also discussed from the view of direct,
indirect and collider searches. 
The last two searches are especially interesting as they may infer the presence of semi-annihilation processes 
and therefore may distinguish between a $\mathbb{Z}_2$ or $\mathbb{Z}_3$ symmetry. 
Finally we summarize and conclude this work in Section~\ref{sec:4}.


\section{The Model}
The model considered here is quite simple. It was proposed and discussed
briefly in ref.~\cite{Ma:2007gq}. 
We introduce two Dirac fermions $\psi_i$ $(i=1,2)$, 
and two scalar bosons $\eta$ and $\chi$ to the SM with
$\mathbb{Z}_3$ symmetry and lepton number as shown in
Tab.~\ref{tab:1}.\footnote{The $\mathbb{Z}_3$ symmetry could be
interpreted as a remnant symmetry of an extra $U(1)$ 
 symmetry as ref.~\cite{Ko:2014nha}.}
The new scalars $\eta$ and $\chi$ are $SU(2)_L$ doublet and singlet,
respectively. 
Note that more than two Dirac fermions are required to generate
at least two non-zero neutrino mass eigenvalues.\footnote{Several components
of scalar $\chi$ may be added to be consistent with neutrino masses
instead of introducing multi-fermions~\cite{Ma:2007gq}.} 
Here we add only two Dirac fermions for minimal particle content. 
The Lagrangian of the new particles is
\begin{eqnarray}
\mathcal{L}_N\hspace{-0.2cm}&=&\hspace{-0.2cm}
\overline{\psi_i}\left(i\partial\hspace{-0.22cm}/-m_i\right)\psi_i
+\left(D_\mu\eta\right)^\dag\left(D^\mu\eta\right)
+\partial_\mu\chi^\dag\partial^\mu\chi\nonumber
\\
&&\hspace{-0.2cm}
+\sum_{i,j}\left(
y_{i\alpha}^\nu\eta\overline{\psi_i}P_LL_{\alpha}
+\frac{y_{ij}^L}{2}\chi\overline{\psi_{i}^{\:c}}P_L\psi_{j}
+\frac{y_{ij}^R}{2}\chi\overline{\psi_{i}^{\:c}}P_R\psi_{j}
+\mathrm{h.c.}
\right),
\end{eqnarray}
where $i,j=1,2$, $\alpha=e,\mu,\tau$ is the flavor index and
$L_\alpha=(\nu_\alpha,\ell_\alpha)^T$ is the left-handed lepton doublet. 
We can take $\psi_i$ in the diagonal base without loss
of generality. 
The gauge and $\mathbb{Z}_3$ invariant renormalizable scalar
potential $\mathcal{V}$ is given by 
\begin{eqnarray}
\mathcal{V}\hspace{-0.2cm}&=&\hspace{-0.2cm}
\mu_{\phi}^2\phi^{\dag}\phi+\mu_{\eta}^2\eta^{\dag}\eta
+\mu_{\chi}^2\chi^{\dag}\chi
+\frac{\lambda_1}{4}\left(\phi^{\dag}\phi\right)^2
+\frac{\lambda_2}{4}\left(\eta^{\dag}\eta\right)^2
+\frac{\lambda_\chi}{4}\left(\chi^{\dag}\chi\right)^2\nonumber\\
&&\!\!\!
+\lambda_3\left(\phi^{\dag}\phi\right)\left(\eta^{\dag}\eta\right)
+\lambda_4\left(\phi^{\dag}\eta\right)\left(\eta^{\dag}\phi\right)
+\lambda_{\phi\chi}\left(\phi^{\dag}\phi\right)\left(\chi^{\dag}\chi\right)
+\lambda_{\eta\chi}\left(\eta^{\dag}\eta\right)\left(\chi^{\dag}\chi\right)
\nonumber\\
&&\!\!\!
+\left(\mu_{\chi}'\left(\phi^{\dag}\eta\right)\chi^{\dag}
+\frac{\mu_\chi''}{3!}\chi^{3}
+\mathrm{h.c.}\right),
\end{eqnarray}
where $\phi$ is the SM Higgs doublet.
The second term in the third line softly breaks the lepton number
conservation and is interpreted as the origin of neutrino masses. 
The phases of $\mu_\chi'$ and $\mu_\chi''$ are absorbed by the field
redefinitions of $\eta$ and $\chi$. 
This scalar potential is basically the same as that in the $\mathbb{Z}_3$ DM model
in ref.~\cite{Belanger:2012vp} except the term $(\phi^\dag\eta)\chi^2$. 
This term is forbidden in our case due to the lepton number non-conservation. 
We assume that the new scalar bosons do not have vacuum expectation
value: $\left<\eta\right>=\left<\chi\right>=0$, otherwise the
$\mathbb{Z}_3$ symmetry which stabilizes the DM candidates breaks down. 
The sufficient conditions in order to get such a vacuum
are given by~\cite{Belanger:2012vp}, 
\begin{eqnarray}
\lambda_1,~\lambda_2,~\lambda_\chi,~\lambda_{\phi\chi},~\lambda_{\eta\chi}
\hspace{-0.2cm}&>&\hspace{-0.2cm}0,
\label{eq:vc1}\\
\lambda_3+\lambda_4\hspace{-0.2cm}&>&\hspace{-0.2cm}0,
\label{eq:vc2}\\
\frac{\mu_\chi''^2}{9\lambda_\chi}
+\frac{\mu_\chi'^2}{\left(\lambda_3+\lambda_4\right)}
\hspace{-0.2cm}&<&\hspace{-0.2cm}\mu_\chi^2.
\label{eq:vc4}
\end{eqnarray}
The parameters $\mu_\phi^2$ and $\lambda_1$ are determined by the vacuum 
expectation value and the mass of the SM Higgs boson,
$\langle\phi\rangle$ ($\approx174$ GeV) and $m_h$, as
$\mu_\phi^2=-m_h^2/2\approx-(89~\mathrm{GeV})^2$ and 
$\lambda_1=m_h^2/\langle\phi\rangle^2\approx0.5$.

\begin{table}[t]
\begin{center}
\begin{tabular}{|c||c|c|c|}\hline
               & $\psi_i$ & $\eta$   & $\chi$\\\hhline{|=#=|=|=|}
$SU(2)$        & $\bm{1}$ & $\bm{2}$ & $\bm{1}$\\\hline
$U(1)_Y$       & $0$      & $1/2$    & $0$\\\hline
$\mathbb{Z}_3$ & $1$      & $1$      & $1$\\\hline
L number       & $1/3$    & $-2/3$   & $-2/3$\\\hline
\end{tabular}
\caption{
Charges of new particles where $\psi_i$ $(i=1,2)$ are Dirac
 fermions, $\eta$ and $\chi$ are 
 scalar bosons. For the other particles in the SM, zero
 charge of $\mathbb{Z}_3$ is assigned.}
\label{tab:1}
\end{center}
\end{table}

After the electroweak symmetry breaking, the neutral component of the SM
Higgs boson $\phi$ can be rewritten as 
$\phi^0=\langle \phi\rangle+h/\sqrt{2}$, and the neutral scalars
$\eta^0$ and $\chi$ mix with each other. A mass splitting between their
real and imaginary parts does not occur so they remain as 
complex scalar particles. 
The mass matrix composed by $\eta^0$ and $\chi$ is given by
\begin{equation}
m_{\eta\chi}^2\equiv\left(
\begin{array}{cc}
m_{11}^2 & |m_{12}|^2\\
|m_{12}|^2 & m_{22}^2
\end{array}
\right)=
\left(
\begin{array}{cc}
\mu_{\eta}^2+\left(\lambda_3+\lambda_4\right)\langle \phi\rangle^2 & 
\mu_\chi'\langle \phi\rangle\\
\mu_\chi'\langle \phi\rangle& \mu_{\chi}^2+\lambda_{\phi\chi}\langle
\phi\rangle^2
\end{array}
\right).
\end{equation}
Then the mass matrix is diagonalized by the rotation matrix
\begin{eqnarray}
\left(
\begin{array}{c}
\eta^0\\
\chi
\end{array}
\right)=\left(
\begin{array}{cc}
\cos\alpha & \sin\alpha\\
-\sin\alpha & \cos\alpha
\end{array}
\right)\left(
\begin{array}{c}
\varphi_H\\
\varphi_L
\end{array}
\right),
\end{eqnarray}
with 
\begin{equation}
\tan{2\alpha}=\frac{2|m_{12}^2|^2}{m_{22}^2-m_{11}^2}
=\frac{2\mu_\chi'\langle
\phi\rangle}{\mu_{\chi}^2-\mu_{\eta}^2
+\left(\lambda_{\phi\chi}-\lambda_3-\lambda_4\right)\langle \phi\rangle^2
},
\end{equation}
where $\varphi_H$ and $\varphi_L$ are respectively the mass eigenstates with the
masses $m_H$ and $m_L$ ($m_H>m_L$). 

There are the following relationships among the parameters: 
\begin{eqnarray}
\mu_{\chi}'\hspace{-0.2cm}&=&\hspace{-0.2cm}-2\left(m_H^2-m_L^2\right)
\frac{\cos\alpha\sin\alpha}{\langle \phi\rangle},
\label{eq:lambda}\\
\mu_\chi^2\hspace{-0.2cm}&=&\hspace{-0.2cm}m_H^2\sin^2\alpha+m_L^2\cos^2\alpha
-\lambda_{\phi\chi}\langle \phi\rangle^2,\\
m_{\eta^+}^2\hspace{-0.2cm}&=&\hspace{-0.2cm}m_L^2\sin^2\alpha+m_H^2\cos^2\alpha
-\lambda_4\langle \phi\rangle^2,
\label{eq:charged_mass}\\
\mu_\eta^2\hspace{-0.2cm}&=&\hspace{-0.2cm}m_L^2\sin^2\alpha+m_H^2\cos^2\alpha
-\left(\lambda_3+\lambda_4\right)\langle \phi\rangle^2,
\label{eq:mu_eta}
\end{eqnarray}
where $m_{\eta^+}$ is the mass of the electromagnetic charged scalar $\eta^+$. 
Thus we can take $m_L^2$, $m_H^2$, $\sin\alpha$, $\lambda_2$, $\lambda_3$,
$\lambda_4$, $\lambda_\chi$, $\lambda_{\phi\chi}$, $\lambda_{\eta\chi}$,
$\mu_\chi''$ as the new independent parameter set in the scalar
potential. 
%

\subsection{Neutrino Mass Matrix}
\begin{figure}[t]
\begin{center}
\includegraphics[scale=1]{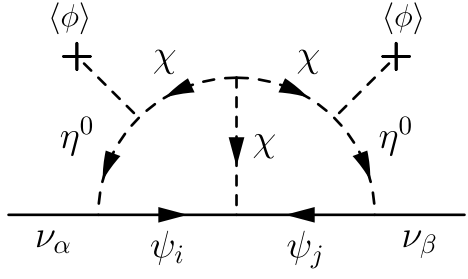}
\caption{Diagram of neutrino mass generation at the two-loop level.}
\label{fig:1}
\end{center}
\end{figure}

In the model, the neutrino masses are induced at the two-loop level as
shown in Fig.~\ref{fig:1}~\cite{Ma:2007gq}. The neutrino mass matrix is calculated as 
\begin{eqnarray}
\left(m_{\nu}\right)_{\alpha\beta}\hspace{-0.2cm}&=&\hspace{-0.2cm}
\sum_{i,j}\frac{y_{i\alpha}^{\nu}y_{j\beta}^{\nu}\sin^22\alpha}{16(4\pi)^4}
\mu_\chi''
\biggl[y_{ij}^L\left(I_{L}\right)_{ij}+y_{ij}^R\left(I_{R}\right)_{ij}\biggr],
\label{eq:neutrino_mass}
\end{eqnarray}
where the loop function $I_R$ is given below: 
\begin{eqnarray}
\left(I_{R}\right)_{ij}\hspace{-0.2cm}&=&\hspace{-0.2cm}
\sin^2\alpha\left(I_{Rij}^{HHH}-I_{Rij}^{HHL}-I_{Rij}^{LHH}+I_{Rij}^{LHL}\right)
\nonumber\\
&&\hspace{-0.2cm}
+\cos^2\alpha\left(I_{Rij}^{HLH}-I_{Rij}^{HLL}-I_{Rij}^{LLH}+I_{Rij}^{LLL}\right),
\end{eqnarray}
and the function $I_{L}$ is obtained by substituting $R\rightarrow L$. 
The functions $I_{Lij}^{abc}$ and $I_{Rij}^{abc}$ are given by 
\begin{eqnarray}
I_{Lij}^{abc}\hspace{-0.2cm}&=&\hspace{-0.2cm}
\frac{m_j}{m_i}\int_0^1\hspace{-0.1cm} dxdydz\frac{\delta(x+y+z-1)}{y(1-y)}
\left[
\frac{\xi_i^a\log\xi_i^a}
{(1-\xi_i^a)(\xi_i^a-\xi_{ij}^{bc})}
-\frac{\xi_{ij}^{bc}\log\xi_{ij}^{bc}}
{(1-\xi_{ij}^{bc})(\xi_{ij}^{bc}-\xi_i^{a})}
\right],\\
I_{Rij}^{abc}\hspace{-0.2cm}&=&\hspace{-0.2cm}
\int_0^1\hspace{-0.1cm} dxdydz\frac{\delta(x+y+z-1)}{1-y}
\left[
\frac{{\xi_i^{a}}^2\log\xi_i^a}
{(1-\xi_i^a)(\xi_i^a-\xi_{ij}^{bc})}
-\frac{{\xi_{ij}^{bc}}^2\log\xi_{ij}^{bc}}
{(1-\xi_{ij}^{bc})(\xi_{ij}^{bc}-\xi_i^{a})}
\right],
\end{eqnarray}
with the parameters $\xi_{ij}^a$ and $\xi_{ij}^{bc}$ defined as 
\begin{equation}
\xi_i^a\equiv\frac{m_a^2}{m_i^2},\quad
\xi_{ij}^{bc}\equiv
\frac{xm_j^2+ym_b^2+zm_c^2}{y(1-y)m_i^2},\quad
a,b,c=H,L,\quad
i,j=1,2.
\label{eq:xi}
\end{equation}

As will be discussed later,
the Yukawa coupling $y^\nu$ should be naively $y^\nu_{i\alpha}\lesssim10^{-2}$ to avoid
the constraint of LFV. If we take $y_{i\alpha}^{\nu}\sim 0.01$,  $\sin\alpha\sim0.1$ and
$(I_L)_{ij}\sim (I_R)_{ij}\sim 0.1$ for example, the required strength of the other couplings
are estimated as $\mu_\chi''\sim10~\mathrm{GeV}$ and $y_{ij}^L\sim y_{ij}^R\sim1$
in order to obtain the appropriate 
neutrino mass scale $m_\nu\sim0.1~\mathrm{eV}$. 
The neutrino mass matrix should be diagonalized by the
Pontecorvo-Maki-Nakagawa-Sakata (PMNS) matrix. 
Although we do not analyze the flavor structure here, it would be
possible to obtain the observed values in the PMNS matrix because of
many parameters in the formula of the neutrino mass matrix.

\subsection{Experimental Constraints}
\begin{figure}[t]
\begin{center}
\includegraphics[scale=1]{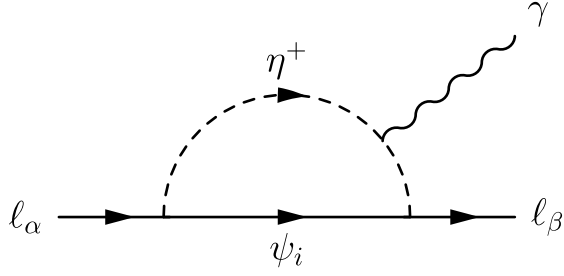}
\caption{
Feynman diagram of LFV process $\ell_\alpha\to\ell_\beta\gamma$.}
\label{fig:meg}
\end{center}
\end{figure}

The LFV process $\ell_\alpha\to\ell_\beta\gamma$ is depicted in Fig.~\ref{fig:meg}.
 The branching ratio for this process~is
\begin{equation}
\mathrm{Br}\left(\ell_{\alpha}\to\ell_{\beta}\gamma\right)
=\frac{3\alpha_{\mathrm{em}}}{64\pi G_F^2}
\left|\sum_{i}\frac{y_{i\beta}^{\nu*}y_{i\alpha}^{\nu}}{m_i^2}
F_i^{\mathrm{loop}}\right|^2
\mathrm{Br}\left(\ell_{\alpha}\to\ell_{\beta}\nu_{\alpha}
\overline{\nu_{\beta}}\right),
\label{eq:lfv}
\end{equation}
where $G_F$ is Fermi constant and the loop function
$F_i^{\mathrm{loop}}$ is given by 
\begin{equation}
F_i^{\mathrm{loop}}=\cos^4\alpha F_2'\left(\xi_i^H,\xi_i^H\right)
+2\cos^2\alpha\sin^2\alpha F_2'\left(\xi_i^H,\xi_i^L\right)
+\sin^4\alpha F_2'\left(\xi_i^L,\xi_i^L\right),
\end{equation}
and $F_2'(x,y)$ is defined as 
\begin{equation}
F_2'\left(x,y\right)=
\frac{f(x)-f(y)}{x-y} \quad\mbox{with}\quad
f(x)=-\frac{5-27x+27x^2-5x^3+6x^2(x-3)\log{x}}{36(1-x)^3}.
\end{equation}
In particular when we take the limit of $y\to x$, the function
$x^{-1}F_2'\left(x^{-1},x^{-1}\right)$ corresponds to 
the function $F_2(x)$
defined in ref.~\cite{Ma:2001mr}. The most stringent constraint comes from
the $\mu\to e\gamma$ whose upper bound of the branting ratio is
$\mathrm{Br}(\mu\to e\gamma)\leq 5.7\times10^{-13}$~\cite{Adam:2013mnn},
and it will be improved to $6\times10^{-14}$ in
future~\cite{Baldini:2013ke}. 
Choosing $F_i^{\mathrm{loop}}\sim0.1$ with $\xi_i^{H}\sim\xi_i^{L}\sim1$, the
requirement for the constraint is roughly written as 
\begin{equation}
\left|y^{\nu*}_{ie}y^\nu_{i\mu}\right|\left(\frac{m_i}{100~\mathrm{GeV}}\right)^{-2}
\lesssim5\times10^{-5}. 
\end{equation}
Hence for example when $m_i=200~\mathrm{GeV}$, the Yukawa coupling is
restricted to $y^{\nu}_{i\alpha}\lesssim10^{-2}$ 
if we do not assume a certain flavor structure.
Another solution to escape the LFV constraint is to assume a specific flavor
structure or a diagonal form for the Yukawa matrix.\footnote{One may confuse
the meaning of ``diagonal'' here  
since the Yukawa matrix $y^\nu$ is not a square matrix. In our case, ``diagonal'' means at
least $y_{1\mu}^\nu=y_{2e}^\nu=y_{1\tau}^\nu=y_{2\tau}^\nu=0$.} 
In this case, $y^\nu$ does not contribute to LFV and 
we can take ${\cal O}(1)$ coupling in some elements of $y^\nu$.
Then the neutrino mixing is derived from
the other Yukawa couplings $y^L$ and $y^R$. 
As we will discuss later,
this solution is interesting because larger $\mathbb{Z}_3$ DM may be more easily detected.
The constraint from another LFV process $\mu\to3e$
might be taken into account depending on the parameter space. 
This process would be enhanced compared to the $\mu\to
e\gamma$ process by the box diagrams when the Yukawa
coupling $y^\nu$ is large enough, 
as has been discussed in
ref.~\cite{Aoki:2011zg, Toma:2013zsa}. 

The mass difference between the charged scalar $\eta^+$ and the neutral
scalars $\varphi_{H,L}$ is constrained by ElectroWeak Precision Tests
(EWPT). Basically, the calculation for our model is the same with the
inert doublet model.
The new contribution to T-parameter, $\Delta T$ is calculated
as~\cite{Barbieri:2006dq}
\begin{equation}
\alpha_{\mathrm{em}}\Delta T=
\frac{1}{2(4\pi)^2\langle\phi\rangle^2}
\biggl[
\cos^2\alpha F(m_{\eta^+}^2,m_H^2)+
\sin^2\alpha F(m_{\eta^+}^2,m_L^2)
\biggr],
\end{equation}
where the function $F(x,y)$ is 
\begin{equation}
F(x,y)=\frac{x+y}{2}-\frac{xy}{x-y}\log\left(\frac{x}{y}\right).
\end{equation}
From these formulae, the constraint on the mass difference between $\eta^+$
and $\varphi_{H,L}$ is approximately given by~\cite{Barbieri:2006dq, Baak:2012kk} 
\begin{equation}
\cos^2\alpha\left(m_{\eta^+}-m_H\right)^2+\sin^2\alpha\left(m_{\eta^+}-m_L\right)^2
\lesssim (140~\mathrm{GeV})^2.
\end{equation}

\section{Dark Matter Properties}
\label{sec:3}
There are two DM candidates in this model, the lightest
Dirac fermion $\psi_1$ or the lightest mass eigenstate of the scalar
boson $\varphi_L$. 
Since the decay process $\psi_1\to\varphi_L\nu_\alpha$ or
$\varphi_L\to\psi_1\nu_\alpha$ is possible in the model
depending on the mass spectrum, either of them can be DM. 
We rename hereafter the Dirac fermion DM as $\psi$ with the mass $m_\psi$
and the scalar one 
as $\varphi$ with the mass $m_\varphi$, and discuss the DM
properties in the following.

\subsection{Dirac Fermion Dark Matter}
\begin{figure}[t]
\begin{center}
\includegraphics[scale=0.8]{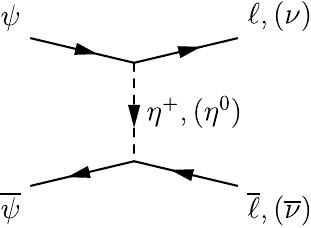}
\hspace{0.2cm}
\includegraphics[scale=0.8]{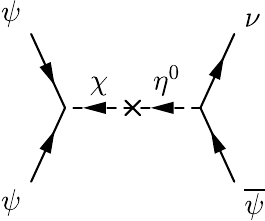}\\
\vspace{0.2cm}
\hspace{1cm}
\includegraphics[scale=0.8]{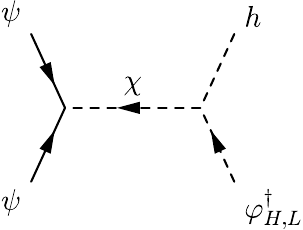}
\hspace{0.2cm}
\includegraphics[scale=0.8]{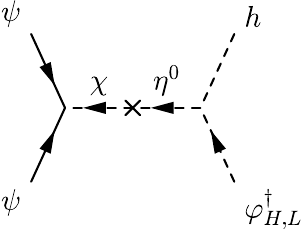}
\hspace{0.2cm}
\includegraphics[scale=0.8]{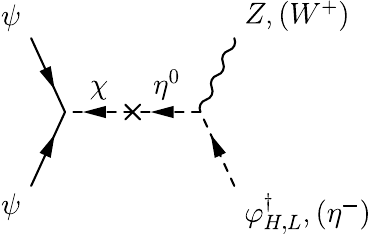}
\caption{Diagrams of (semi-)annihilation processes of Dirac DM
 $\psi$. The t-channel diagram of the top right one also exists.}
\label{fig:2-body}
\end{center}
\end{figure}

For the Dirac fermion DM $\psi$,
in addition to the annihilation channels, we have some semi-annihilation channels
like $\psi\psi\to\nu\overline{\psi}, h\varphi^\dag, Z\varphi^\dag$ as shown in
Fig.~\ref{fig:2-body}. 
The evolution of the number density of Dirac DM is determined by
the Boltzmann equation:
\begin{equation}
\frac{dn_\psi}{dt}+3Hn_\psi=
-\langle\sigma{v}_{\psi\overline{\psi}}\rangle
\left(n_\psi^2-{n_{\psi}^{\mathrm{eq}}}^2\right)
-\frac{1}{2}\langle\sigma{v}_{\psi\psi}\rangle
\left(n_\psi^2-n_{\psi}n_{\psi}^{\mathrm{eq}}\right),
\label{Boltzmann}
\end{equation}
where the number densities of $\psi$ and $\overline{\psi}$ are assumed
to be the same $n_\psi=n_{\overline{\psi}}$. This assumption is valid as
long as CP invariance is considered. 
The first term in the right-hand side of Eq.~(\ref{Boltzmann}) implies
the standard annihilation processes and the second 
term corresponds to the semi-annihilation processes. 
The factor $1/2$ in the second term
comes from taking into account the processes
$\psi\psi\to X\overline{\psi}$ and
$\overline{\psi}\overline{\psi}\to\overline{X}\psi$ both where $X$ is
any set of the SM particles. 
The contribution of the decay processes such as $\varphi_L\to\psi\nu_\alpha$
is negligible unless the decaying particle has an extremely long lifetime,
considered since the DM is still in the thermal bath when the decay 
process decouples. 
We construct our own model with
LanHEP~\cite{Semenov:2008jy}\footnote{Feynrules is useful to
construct a model~\cite{Alloul:2013bka}.}, and then
micrOMEGAs is used to solve the Boltzmann
equation numerically~\cite{Belanger:2013oya}. 
In general, off-shell annihilation processes such as $\psi\psi\to
Z^*\varphi_{H,L}^\dag$ are also included in our numerical analysis. 

If any semi-annihilation channels are not significant, only annihilation processes
$\psi\overline{\psi}\to\ell\overline{\ell},\nu\overline{\nu}$ affect the thermal relic
density of the Dirac DM. The cross section is 
proportional to $\left|y_{1\alpha}^{\nu*} y_{1\beta}^\nu\right|^2$, 
and the required order of the Yukawa coupling is
roughly 
$y_{1\alpha}^\nu\sim\mathcal{O}(0.1 - 1)$ 
to be compatible
with the observed relic density $\Omega h^2\approx0.12$~\cite{Ade:2013zuv}. 
However the Yukawa coupling is severely
constrained by the LFV processes as have discussed in the previous section. 
One needs roughly $y_{i\alpha}^\nu\lesssim0.01$ to avoid the $\mu\to e\gamma$
constraint. 
Thus it seems to be difficult to satisfy the correct thermal relic
density only with the annihilation channel. 
The effect of semi-annihilation is important to give consistency to both
the DM relic density and LFV as will be discussed in 
Sec.~\ref{sec:small_yukawa}. 
Another solution is the consideration of mass degeneration with the other new
particles~\cite{Suematsu:2009ww, Schmidt:2012yg}, then the relic density
is reduced by co-annihilation with the degenerated particles. In
particular co-annihilation with $\eta$ gives a large contribution to the
effective cross section since it has gauge interactions. 

Meanwhile, taking a larger Yukawa coupling $y^{\nu}$ can be possible if a
special flavor texture to suppress the LFV processes is assumed~\cite{Suematsu:2009ww,
Schmidt:2012yg}. 
This possibility will be discussed in Sec.~\ref{sec:large_yukawa}.

\subsubsection{Small Yukawa Coupling}
\label{sec:small_yukawa}
\begin{table}[t]
\begin{center}
\begin{tabular}{|c||c|c|c|c|c|}\hline
 & $\lambda_{\phi\chi}$ & $m_{L}~[\mathrm{GeV}]$ &
 $m_{H}~[\mathrm{GeV}]$ & $y^\nu$ & $y^L,~y^R$\\
\hhline{|=#=|=|=|=|=|} 
 BM-F1 & $0.1$ & $400$ & $500$ & \multirow{4}{*}{
 $\left(
\begin{array}{ccc}
0.01 & 0.01 & 0.01\\
0.01 & 0.01 & 0.01
\end{array}
\right)$
 } &
		     \multirow{4}{*}{$\mathcal{O}(1)$}\\\cline{1-4}
 BM-F2 & $0.1$ & $400$ & $400$ & &\\\cline{1-4}
 BM-F3 & $1.0$ & $400$ & $500$ & &\\\cline{1-4}
 BM-F4 & $1.0$ & $400$ & $400$ & &\\
 \hhline{|=#=|=|=|=|=|}
 BM-F1$'$ & $0.1$ & $400$ & $500$ & \multirow{4}{*}
{
$
\left(
\begin{array}{ccc}
0.5 & 0 & 0\\
0 & 0.5 & 0
\end{array}
\right)$
}
 &
\multirow{4}{*}{$\mathcal{O}(0.1)$}\\\cline{1-4} 
 BM-F2$'$ & $0.1$ & $400$ & $400$ & &\\\cline{1-4}
 BM-F3$'$ & $1.0$ & $400$ & $500$ & &\\\cline{1-4}
 BM-F4$'$ & $1.0$ & $400$ & $400$ & &\\\hline
\end{tabular}
\end{center}
\caption{
Benchmark parameter sets for Fig.~\ref{fig:omega_f1} and
 \ref{fig:omega_f2}. The other parameters are fixed to
 $\lambda_2=\lambda_\chi=\lambda_{\eta\chi}=0.1$, 
 $\lambda_3=\lambda_4=0.5$ and $m_2=1~\mathrm{TeV}$. 
 The parameter $\mu_\chi''\sin^2\alpha$ is fixed to
 $100~\mathrm{MeV}$ for the upper four sets, and 
 $1~\mathrm{MeV}$ for the lower four sets. 
 The mass of $\eta^+$ is determined by Eq.~(\ref{eq:charged_mass}).}
\label{tab:bm}
\end{table}

\begin{figure}[t]
\begin{center}
\includegraphics[scale=0.6]{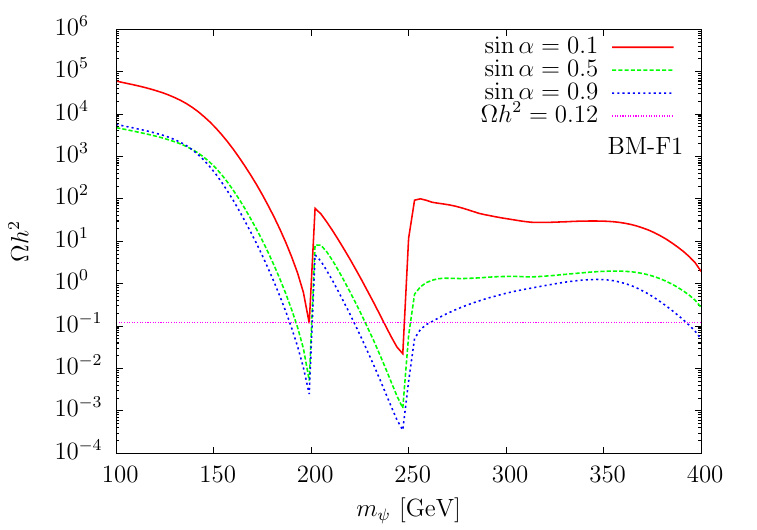}
\includegraphics[scale=0.6]{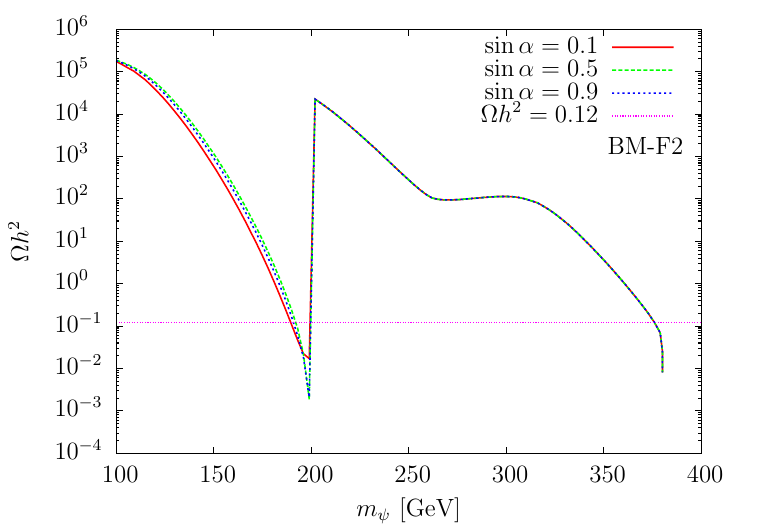}\\
\includegraphics[scale=0.6]{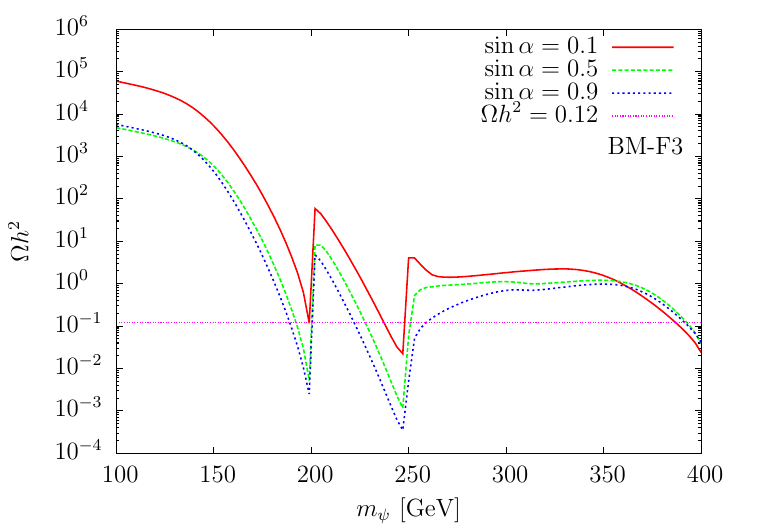}
\includegraphics[scale=0.6]{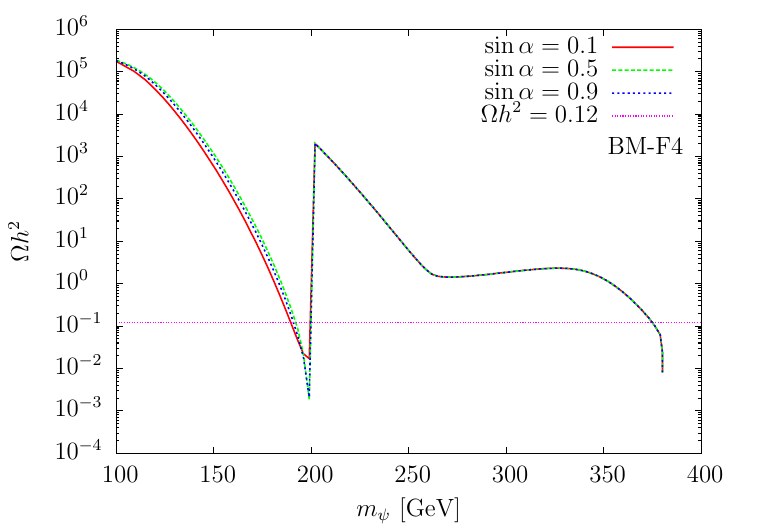}
\caption{DM mass dependence of the relic density. The upper panels are for
 BM-F1 and BM-F2, and the lower panels are for BM-F3 and BM-F4.
}
\label{fig:omega_f1}
\end{center}
\end{figure}

The DM mass dependence of the relic density is illustrated in
Fig.~\ref{fig:omega_f1} for the small Yukawa couplings.
The masses of the $\mathbb{Z}_3$ charged scalars and the other
parameters in the model are fixed to the benchmark sets as shown in
Tab.~\ref{tab:bm}. 
For the parameter sets of BM-F1 and BM-F3, the masses of scalars are set
as $m_{L}=400~\mathrm{GeV}$ and $m_{H}=500~\mathrm{GeV}$, and the
charged scalar mass is fixed by Eq.~(\ref{eq:charged_mass}). 
There are two damped regions, around $200~\mathrm{GeV}$ and $250~\mathrm{GeV}$ in
the left upper and lower panels of Fig.~\ref{fig:omega_f1}. 
The first damped region results from the resonance of $\varphi_L$ in the
semi-annihilation processes $\psi\psi\to\varphi_L\to\nu\overline{\psi}$,
while the second one corresponds to the $\varphi_H$ resonance in mainly
$\psi\psi\to\varphi_H\to W^+\eta^-,Z\varphi_L^\dag$. 
Thus one can satisfy the correct relic density of DM without any
contradictions thanks to the resonances of the semi-annihilation channels.
The co-annihilation with $\varphi_L$ is
effective around $400~\mathrm{GeV}$ for BM-F1 and BM-F3.
When a large $\lambda_{\phi\chi}$ is taken as BM-F3, the
semi-annihilation process $\psi\psi\to\varphi_L\to
h\varphi_L^\dag$ is enhanced even if the mixing $\sin\alpha$ is zero. This
process starts to be effective around $m_{\psi}=(m_h+m_{\varphi_L})/2\approx
260~\mathrm{GeV}$. As found in the lower left panel in
Fig.~\ref{fig:omega_f1}, the relic abundance of DM 
is much reduced in the region of $m_\psi\gtrsim260~\mathrm{GeV}$ 
for large $\lambda_{\phi\chi}$. 
The effect of $\lambda_{\phi\chi}$ is significant in particular for small
mixing angle.

In the right panels for BM-F2 and BM-F4, the masses of $\varphi_L$ and
$\varphi_H$ are both fixed to $400~\mathrm{GeV}$. The behavior is quite
different from the former case. The sharp damping around $200~\mathrm{GeV}$
comes from the dominant semi-annihilation $\psi\psi\to\varphi_{H,L}\to
\nu\overline{\psi}$, but the second damped region disappears. Thus the observed
relic density is achieved only at the 
damped and the co-annihilation region with $\varphi_{H,L}$. 
Another difference from the left panels is that the dependence of
the mixing angle is 
extremely small. This is because when the masses are degenerated, the 
scalar couplings in the potential have almost no dependence on the mixing
angle as one can see from Eq.~(\ref{eq:lambda})$-$(\ref{eq:mu_eta}). 
The maximum DM mass is bounded by the charged scalar mass
$m_{\eta^+}\approx380~\mathrm{GeV}$.

\subsubsection{Large Yukawa Coupling}
\label{sec:large_yukawa}
\begin{figure}[t]
\begin{center}
\includegraphics[scale=0.6]{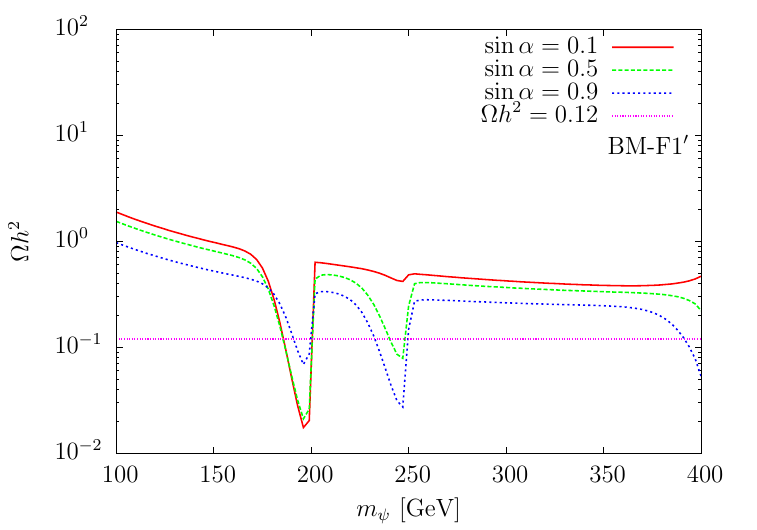}
\includegraphics[scale=0.6]{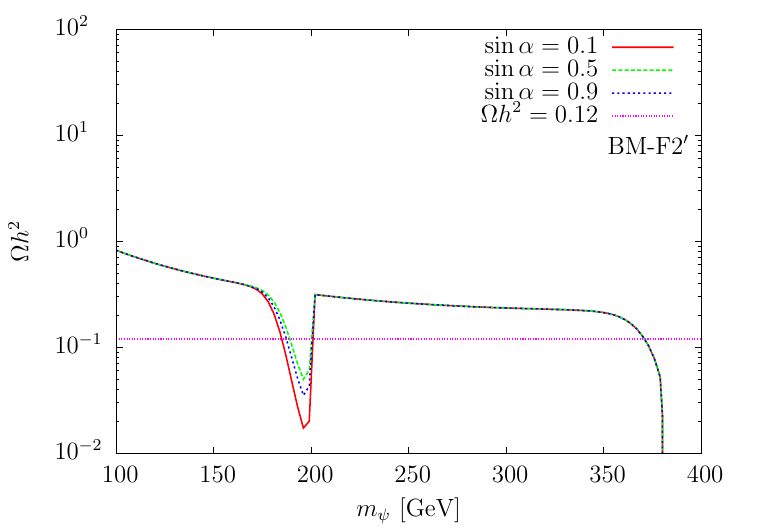}\\
\includegraphics[scale=0.6]{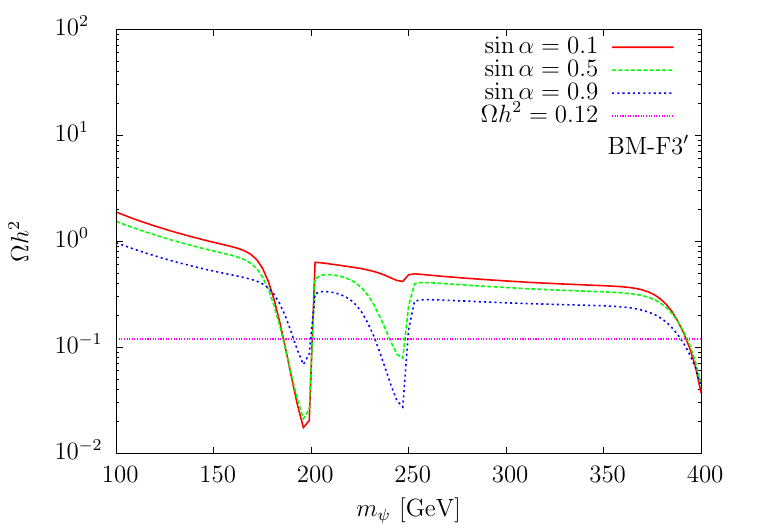}
\includegraphics[scale=0.6]{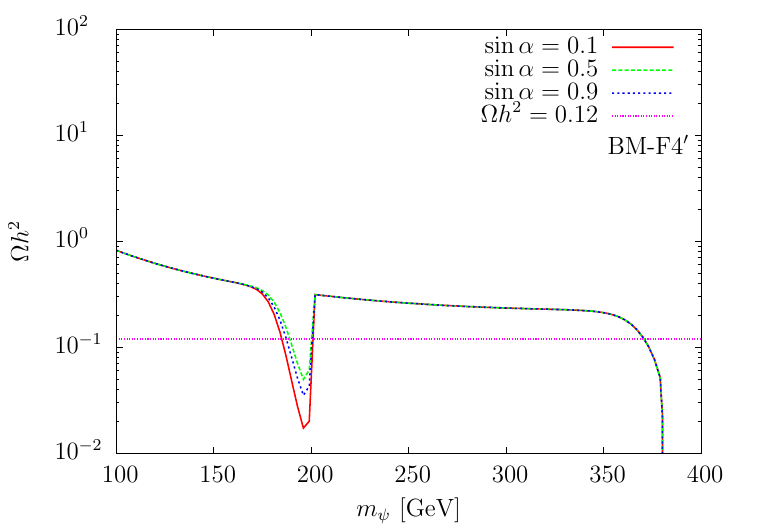}
\caption{DM mass dependence of the relic density for some parameter
 sets. The upper panels correspond to BM-F1$'$ and BM-F2$'$, and the lower
 panels are for BM-F3$'$ and BM-F4$'$. Details of the parameter sets are
 in Tab.~\ref{tab:bm}. 
}
\label{fig:omega_f2}
\end{center}
\end{figure}

Suppressing the LFV while having large Yukawa couplings is allowed
by choosing a specific flavor structure for the Yukawa matrix $y^\nu$.
This case is quite interesting from view of the detectability of the Dirac
DM as we will discuss later.  
In Fig.~\ref{fig:omega_f2}, we show the DM mass dependence of the relic
density for the benchmark sets BM-F1$'$, BM-F2$'$, BM-F3$'$ and
BM-F4$'$. These parameter sets are the same as in Tab.~\ref{tab:bm} for BM-F1,
BM-F2, BM-F3 and BM-F4 respectively, 
but the values of the parameters $y^\nu$, $y^L$, $y^R$ and $\mu_\chi''$
are different. The Yukawa coupling $y^\nu$ is
fixed to $y_{1e}^\nu=y_{2\mu}^\nu=0.5$ and $y_{1\mu}^\nu=y_{2e}^\nu=0$
to suppress the $\mu\to e\gamma$ process. The other components of $y^\nu$ are also fixed
adequately. The Yukawa couplings $y^L$ and $y^R$ are taken as 
$y^L\sim y^R\sim \mathcal{O}(0.1)$. 
Then the known neutrino mass scale can be obtained by
adjusting the cubic coupling $\mu_\chi''$ to satisfy 
$\mu_\chi''\sin^2\alpha\sim1~\mathrm{MeV}$ from Eq.~(\ref{eq:neutrino_mass}).

Since the Yukawa coupling $y^\nu$ is large, the standard annihilation channels
$\psi\overline{\psi}\to\ell\overline{\ell},\nu\overline{\nu}$ 
give some contribution to the total annihilation cross section
throughout Fig.~\ref{fig:omega_f2}. 
In these parameter sets, the cross sections of the annihilation and
semi-annihilation processes can be comparable, and 
the dependence of the scalar coupling $\lambda_{\phi\chi}$ becomes
relatively smaller than the case of small Yukawa coupling. This can be
seen in the figure. 
The variation of the relic density in terms of the DM mass for the large
Yukawa coupling is much milder than that for the small Yukawa coupling, as can be
seen from Fig.~\ref{fig:omega_f1} and Fig.~\ref{fig:omega_f2}.

\subsection{Detectability of Dirac Fermion Dark Matter}

\subsubsection{Direct Search}
\begin{figure}[t]
\begin{center}
\includegraphics[scale=1]{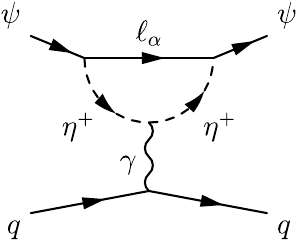}
\hspace{0.2cm}
\includegraphics[scale=1]{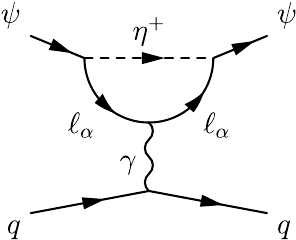}
\hspace{0.2cm}
\includegraphics[scale=1]{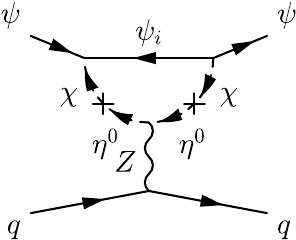}
\caption{Diagram of primary contributions to spin independent and spin dependent cross
 section for Dirac DM $\psi$.}
\label{fig:dd}
\end{center}
\end{figure}

The Dirac DM $\psi$ can interact with quarks at one-loop level via
photon and $Z$ exchanges as depicted in Fig.~\ref{fig:dd}. 
Since the one-loop interactions are described by the Yukawa coupling $y^\nu$, $y^L$
and $y^R$, the case of large Yukawa coupling discussed
above should especially be compared with experimental limits. 
When the Yukawa couplings are small, the scattering cross section is small and will not be
detected in the near future. 
The following relevant interactions are obtained through
the one-loop diagram, 
\begin{eqnarray}
\mathcal{L}_\mathrm{eff}\hspace{-0.2cm}&=&\hspace{-0.2cm}
a_\psi\overline{\psi}\gamma^\mu\psi\partial^{\nu}F_{\mu\nu}
+\left(\frac{\mu_\psi}{2}\right)\overline{\psi}\sigma^{\mu\nu}\psi
F_{\mu\nu}
+c_\psi A_\mu\overline{\psi}\gamma^\mu\psi\nonumber\\
&&\hspace{-0.2cm}
+Z_\mu\overline{\psi}\left(V_\psi\gamma^\mu+A_\psi\gamma^\mu\gamma_5\right)\psi,
\end{eqnarray}
where $F_{\mu\nu}$ is the electromagnetic field strength and the
couplings $a_\psi$, $\mu_\psi$, $c_\psi$ are given by~\cite{Schmidt:2012yg} 
\begin{eqnarray}
a_\psi\hspace{-0.2cm}&=&\hspace{-0.2cm}
-\sum_{\alpha}\frac{\left|y_{1\alpha}^\nu\right|^2e}{4(4\pi)^2m_{\eta^+}^2}
I_\mathrm{a}\left(\frac{m_\psi^2}{m_{\eta^+}^2},\frac{m_\alpha^2}{m_{\eta^+}^2}\right),\\
\mu_\psi\hspace{-0.2cm}&=&\hspace{-0.2cm}
-\sum_{\alpha}\frac{\left|y_{1\alpha}^\nu\right|^2e}{4(4\pi)^2m_{\eta^+}^2}2m_\psi
I_\mathrm{m}\left(\frac{m_\psi^2}{m_{\eta^+}^2},\frac{m_\alpha^2}{m_{\eta^+}^2}\right),\\
c_\psi\hspace{-0.2cm}&=&\hspace{-0.2cm}
+\sum_{\alpha}\frac{\left|y_{1\alpha}^\nu\right|^2e}{4(4\pi)^2m_{\eta^+}^2}q^2
I_\mathrm{c}\left(\frac{m_\psi^2}{m_{\eta^+}^2},\frac{m_\alpha^2}{m_{\eta^+}^2}\right),
\end{eqnarray}
where $q^2$ is the transfer momentum to the gauge boson.
The loop functions $I_\mathrm{a}(x,y)$, $I_\mathrm{m}(x,y)$ and 
$I_\mathrm{c}(x,y)$ are given in ref.~\cite{Schmidt:2012yg}. 
The factor $\log{y}$ is included in the loop
function $I_\mathrm{a}(x,y)$, and it leads to some enhancement of the interaction. 
The interactions with $Z$ boson are calculated as 
\begin{eqnarray}
V_\psi\hspace{-0.2cm}&=&\hspace{-0.2cm}
-\frac{g_2\sin^22\alpha}
{32\left(4\pi\right)^2\cos\theta_W}
\sum_{i}
\left(\left|y^L_{i1}\right|^2+\left|y^R_{i1}\right|^2\right)
\sum_{a,b=H,L}\mathrm{sgn}(a,b)I_{V1}\left(\xi^i,\xi^a,\xi^b\right)\nonumber\\
&&\hspace{-0.2cm}
-\frac{g_2\sin^22\alpha}
{32\left(4\pi\right)^2\cos\theta_W}
\sum_{i}
\Bigl(y_{i1}^L{y_{i1}^R}^*+{y_{i1}^L}^*y_{i1}^R\Bigr)
\sum_{a,b=H,L}\mathrm{sgn}(a,b)I_{V2}\left(\xi^i,\xi^a,\xi^b\right),\\
A_\psi\hspace{-0.2cm}&=&\hspace{-0.2cm}
+\frac{g_2\sin^22\alpha}
{32\left(4\pi\right)^2\cos\theta_W}
\sum_{i}
\left(\left|y^L_{i1}\right|^2-\left|y^R_{i1}\right|^2\right)
\sum_{a,b=H,L}\mathrm{sgn}(a,b)I_A\left(\xi^i,\xi^a,\xi^b\right),
\end{eqnarray}
with
\begin{equation}
\xi^i\equiv\frac{m_i^2}{m_\psi^2},\qquad
\xi^L\equiv\frac{m_L^2}{m_\psi^2},\qquad
\xi^H\equiv\frac{m_H^2}{m_\psi^2}.
\end{equation}
The coefficient $g_2$ is the $SU(2)_L$ gauge coupling constant and
$\sin\theta_W$ is the Weinberg angle. 
The sign function $\mathrm{sgn}(a,b)$ is defined as 
$\mathrm{sgn}(H,H)=\mathrm{sgn}(L,L)=1$ and
$\mathrm{sgn}(H,L)=\mathrm{sgn}(L,H)=-1$, and the loop functions
$I_{V1}$, $I_{V2}$ and $I_A$ are
\begin{eqnarray}
I_{V1}\left(x,y,z\right)\hspace{-0.2cm}&=&\hspace{-0.2cm}
\int_0^1\frac{1-2u+2u^2}{2\left(y-z\right)}
\log\left(\frac{ux+(1-u)(y-u)}{ux+(1-u)(z-u)}\right)du,\\
I_{V2}\left(x,y,z\right)\hspace{-0.2cm}&=&\hspace{-0.2cm}
\int_0^1\frac{\sqrt{x}}{y-z}
\log\left(\frac{ux+(1-u)(y-u)}{ux+(1-u)(z-u)}\right)du,\\
I_A\left(x,y,z\right)\hspace{-0.2cm}&=&\hspace{-0.2cm}
\int_0^1\frac{1}{2\left(y-z\right)}
\log\left(\frac{ux+(1-u)(y-u)}{ux+(1-u)(z-u)}\right)du.
\end{eqnarray}
The couplings $V_\psi$ and $A_\psi$ vanish when the $\mathbb{Z}_3$ charged bosons
$\varphi_L$ and $\varphi_H$ are completely degenerate. 
Since these interactions are proportional to $\sin^22\alpha$, one can
expect that a moderate cross section is derived when the mixing angle is
large. 

In these effective interactions, the couplings $a_\psi$, $c_\psi$ and $V_\psi$
contribute to the four Fermi vector interaction with quarks:
$b_\psi\overline{q}\gamma_\mu q\overline{\psi}\gamma^{\mu}\psi$. 
The contribution via $Z$ boson is small compared to the photon
contribution if the Yukawa couplings $y^\nu$ and $y^{L,R}$ are of the same order. 
Neglecting the $Z$ boson contribution, the four Fermi interaction
$b_\psi$ is given by $b_\psi=(a_\psi+c_\psi/q^2)e$, namely 
\begin{equation}
b_\psi=-\sum_{\alpha}\frac{|y_{1\alpha}^\nu|^2e^2}{4(4\pi)^2m_{\eta^+}^2}
I_\mathrm{b}\left(\frac{m_\psi^2}{m_{\eta^+}^2},\frac{m_\alpha^2}{m_{\eta^+}^2}\right),
\end{equation}
where $I_\mathrm{b}(x,y)=I_\mathrm{a}(x,y)-I_\mathrm{c}(x,y)$. 
The spin independent elastic cross section with nucleus at zero transfer
momentum via the vector interaction is calculated as 
\begin{equation}
\sigma_{\mathrm{SI}}=Z^2\frac{b_\psi^2}{\pi}\frac{m_\psi^2m_A^2}{\left(m_\psi+m_A\right)^2}, 
\end{equation}
where $Z$ and $m_A$ are atomic number and mass of
nucleus. For example, the elastic cross section to scatter with a
proton with the parametrization BM-F1$'$ is estimated as
$\sigma_{\mathrm{SI}}\sim10^{-45}~\mathrm{cm^2}$, which is slightly
below the upper bound from LUX for
$m_\psi\gtrsim100~\mathrm{GeV}$~\cite{Akerib:2013tjd}. 
Thus it would be detected by near-future direct detection
experiments such XENON1T~\cite{Aprile:2012zx}. 
In fact there is another contribution to the spin independent cross section
via the magnetic moment of DM 
$\mu_\psi$. However this contribution diverges at zero recoil energy and we
cannot define adequately the total cross section at zero momentum
transfer. Moreover this contribution is sub-dominant
because of the enhancement factor $\log{y}$ of $I_\mathrm{a}(x,y)$ in
the effective coupling $a_\psi$, as already pointed out 
above. Thus we do not include this contribution in our
discussion. If more careful treatment is required, this contribution
should be taken into account.

The couplings $\mu_\psi$ and $A_\psi$ contribute to the spin dependent cross
section. As with the the spin independent cross section, the effective
interaction via a $Z$ boson is sub-dominant. Thus the spin dependent cross
section at zero momentum transfer is simply given by 
\begin{equation}
\sigma_{\mathrm{SD}}=
\frac{2\mu_\psi^2\mu_A^2}{\pi}\frac{m_\psi^2m_A^2}{\left(m_\psi+m_A\right)^2}
\left(\frac{J_A+1}{3J_A}\right),
\end{equation}
where $\mu_A$ and $J_A$ are the magnetic moment and the spin of nucleus,
respectively. For the large Yukawa benchmark parameter sets, the order
of the cross section with a proton is roughly estimated as
$\sigma_{\mathrm{SD}}\sim10^{-45}~\mathrm{cm^2}$. 
The present strongest upper bound on the spin dependent cross section
is given as $\sigma_{\mathrm{SD}}\lesssim10^{-39}~\mathrm{cm^2}$ by
COUPP~\cite{Behnke:2012ys} and SIMPLE\cite{Felizardo:2011uw}, which is 
too weak to constrain the model. 

In addition, there are further more severe constraints on the spin dependent cross section
 from the search for neutrinos from the
Sun by IceCube~\cite{Aartsen:2012kia}. 
These limits hold when 
the capture rate and the annihilation rate of DM in the Sun are in equilibrium.
As a result, the capture rate which depends on both the spin
independent and dependent cross section is constrained, depending on
annihilation mode of DM.

\subsubsection{Indirect Search}
\begin{figure}[t]
\begin{center}
\includegraphics[scale=0.7]{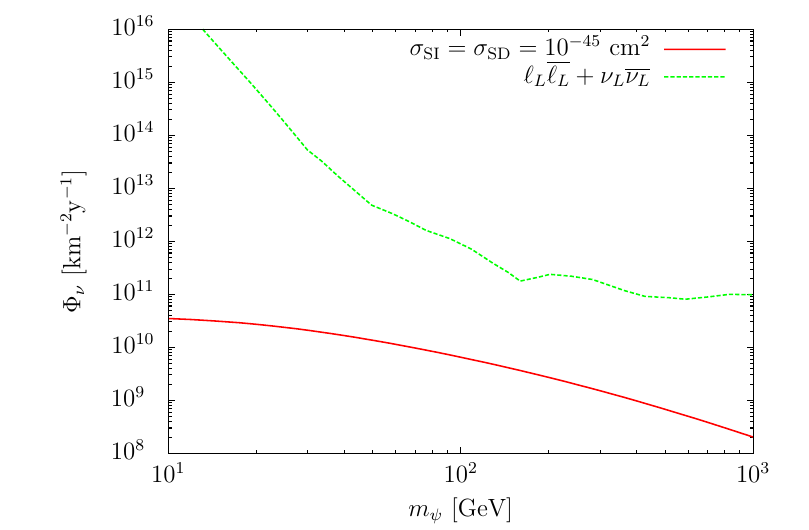}
\caption{
DM mass dependence of total neutrino flux $\Phi_\nu$. 
The expected neutrino flux in the model is given by the red solid line
 where SI and SD cross sections are fixed to
 $\sigma_{\mathrm{SI}}=\sigma_{\mathrm{SD}}=10^{-45}~\mathrm{cm^2}$. 
 The upper bound for the channel
 $\ell_L\overline{\ell_L}+\nu_L\overline{\nu_L}$ obtained from the 
 IceCube data analysis~\cite{Ibarra:2014vya} is given by the green dotted
 line.}
\label{fig:nu_flux}
\end{center}
\end{figure}

In general, semi-annihilation processes are present when we consider a larger symmetry than $\mathbb{Z}_2$
for stabilizing DM,
such as a $\mathbb{Z}_3$ symmetry. A characteristic implication of
semi-annihilation may be observed in indirect searches of
DM~\cite{D'Eramo:2010ep, D'Eramo:2012rr, Aoki:2012ub}. 
In our case, we have two channels for generating monochromatic neutrinos,
$\psi\overline{\psi}\to \nu\overline{\nu}$ and
$\psi\psi\to\nu\overline{\psi}$, whose energies are determined
kinematically as $E_{\nu}=m_\psi$ and $3m_\psi/4$ respectively. 
Thus a double peak may be detected in the neutrino flux from the
galaxy or the Sun as a signature of the model with the semi-annihilation of
DM~\cite{D'Eramo:2010ep}. 
A large Yukawa coupling $y^\nu$ with a special flavor
structure is necessary to see the signal of the double peak. 

For the large Yukawa parameter sets, the standard annihilation
$\psi\overline{\psi}\to \nu\overline{\nu}$ and the semi-annihilation $\psi\psi\to
\nu\overline{\psi}$ can be the main channels and comparable each other
in some parameter regions, as shown in the previous section. 
The present upper bound for the annihilation
cross section into neutrinos from the galactic center is
$\langle\sigma{v}\rangle_{\nu\overline{\nu}}
\lesssim10^{-22}~\mathrm{cm^3/s}$~\cite{Abbasi:2012ws}, which is far
from the canonical annihilation cross section of the thermal DM. 
The neutrinos from the Sun give a somewhat stronger bound. 
Monochromatic neutrino emission from DM annihilation has been studied in
ref.~\cite{Barger:2007hj, Delaunay:2008pc, Andreas:2009hj, Farzan:2011ck}.
The differential neutrino flux from the Sun is calculated as
\begin{equation}
\frac{d\Phi_\nu}{dE_\nu}=
\frac{1}{2}\frac{C_\odot}{4\pi d^2}
\left[
2\mathrm{Br}(\psi\overline{\psi}\to\nu\overline{\nu})
\frac{dN_{\nu\overline{\nu}}}{dE_\nu}+
\mathrm{Br}(\psi\psi\to\nu\overline{\psi})
\frac{dN_{\nu\overline{\psi}}}{dE_\nu}
\right],
\label{eq:nu_f}
\end{equation}
where 
the factor $2$ of the first term in Eq.~(\ref{eq:nu_f}) comes from two
neutrinos in the final state, 
$d=1.49\times10^{8}~\mathrm{km}$ is the distance between the Earth and
Sun, $\mathrm{Br}$ is the branching ratio of the process,
$dN_{\nu\overline{\nu}}/dE_\nu$ and $dN_{\nu\overline{\psi}}/dE_\nu$ are
the energy spectra of neutrino for each channel and they are simply written
by the delta function in our case. 
The capture rate in the Sun $C_\odot$ is estimated
by using micrOMEGAs~\cite{Belanger:2013oya}. 
In particular for $100~\mathrm{GeV}\lesssim
m_\psi\lesssim1~\mathrm{TeV}$, the capture rate is simply evaluated as
\begin{equation}
C_{\odot}\approx
\left(
\frac{1.2\times10^{20}}{\mathrm{s}}
\right)
\left[
\left(\frac{100~\mathrm{GeV}}{m_\psi}\right)^{1.7}
\left(\frac{\sigma_{\mathrm{SI}}}{10^{-45}~\mathrm{cm^2}}\right)
+
\left(\frac{100~\mathrm{GeV}}{m_\psi}\right)^{1.9}
\left(\frac{\sigma_{\mathrm{SD}}}{10^{-42}~\mathrm{cm^2}}\right)
\right],
\end{equation}
where the Maxwell-Boltzmann distribution is assumed for the DM velocity
distribution function with the dispersion $v_0=270~\mathrm{km/s}$ and
the local DM density $\rho_\odot=0.3~\mathrm{GeV/cm^3}$. 
The total neutrino flux is simply calculated as 
\begin{equation}
\Phi_\nu=
\frac{1}{2}\frac{C_\odot}{4\pi d^2}
\Bigl(2\mathrm{Br}(\psi\overline{\psi}\to\nu\overline{\nu})
+\mathrm{Br}(\psi\psi\to\nu\overline{\psi})\Bigr)~[\mathrm{km^{-2}y^{-1}}].
\label{eq:nu_flux}
\end{equation}
The DM mass dependence of the total neutrino flux is shown in
Fig.~\ref{fig:nu_flux} where both $\sigma_{\mathrm{SI}}$ and 
$\sigma_{\mathrm{SD}}$ are fixed to $10^{-45}~\mathrm{cm^2}$, as we evaluated for the
large Yukawa parameter sets, and the sum of the branching
ratios in Eq.~(\ref{eq:nu_flux}) is taken as $1$. 
In Fig.~\ref{fig:nu_flux}, the IceCube upper bound for
$\ell_L\overline{\ell_L}+\nu_L\overline{\nu_L}$ annihilation channel is
also shown. The IceCube bound is obtained by
converting the upper bound on the elastic cross section, which has been calculated
from the IceCube data~\cite{Ibarra:2014vya}.
This bound should be understood as a rough reference limit since 
the annihilation channel is not exactly the same as with our monochromatic case. 
One finds that the expected neutrino flux in the model is two orders of
magnitude smaller than the upper bound at most. 
To more thoroughly compare the predicted flux with experiments, 
the effects of neutrino oscillation and propagation together with
experimental details should be taken into account. 

For the case of small Yukawa
coupling, only a single peak of monochromatic neutrino from the
semi-annihilation can be seen. It would be difficult to distinguish this
from the monochromatic neutrino from the annihilation of typical
$\mathbb{Z}_2$ symmetric DM.

\subsubsection{Collider Prospects}
Generally, DM with a $\mathbb{Z}_3$ symmetry will give different collider
signatures from $\mathbb{Z}_2$ DM~\cite{Agashe:2010gt, Agashe:2010tu}. 
While only one DM is generated in the final state from the decay of the
$\mathbb{Z}_2$ mother particle, one or two DM particles are produced in the
decay of the $\mathbb{Z}_3$ charged particle. 
Therefore in the $\mathbb{Z}_3$ symmetric model, for instance, if 
the signals of one DM and two DM in the final state have the same visible particles 
and the intermediate particles are off-shell, the double-kinematic
edge would be seen in the invariant mass distribution of the visible
particles as a prospect of $\mathbb{Z}_3$ DM. In our model, however,
such decay channels are not expected.

On the other hand, if the intermediate particles are on-shell,
the invariant mass distribution will have a different shape in 
$\mathbb{Z}_2$ and $\mathbb{Z}_3$ symmetric models~\cite{Agashe:2010gt, Agashe:2010tu}.
Considering the decay channel of $\mathbb{Z}_3$ charged particle
with two visible particles separated by a DM, the invariant mass
distribution for two visible particles has a cusp.
Since such a cuspy feature cannot be present in the decay of
$\mathbb{Z}_2$ odd particle, it gives one possible way to
discriminate between $\mathbb{Z}_2$ and $\mathbb{Z}_3$ symmetric
models.
In our model, we can consider the decay of 
$\eta^\pm$ which can be produced in pairs via the Drell-Yan process
of $\gamma$ and $Z$ exchange
or singly produced via $W^\pm$ exchange at the Large Hadron Collider.
A concrete example of above decay channel is 
$\eta^+\to\mu^+\psi_2\to\mu^+\overline{\psi}\varphi_H^\dag\to\mu^+\overline{\psi}Z\varphi_L^\dag$,
where the intermediate particles are all on-shell by assuming the mass
hierarchy $m_{\eta^+} > m_2 > m_H>m_L$ and suitable mass differences.
The anti-DM particle $\overline{\psi}$ produced by the decay of $\psi_2$
separates $\mu^+$ and $Z$, which gives rise to a cusp in
the invariant mass distribution for the $\mu^+ Z$ system.
It is noted that the DM mass in this example has to be smaller than
$\sim$ 70 GeV, otherwise the mass difference between $\eta^+$ and
$\varphi_{H,L}$ becomes inconsistent with the EWPT.
For example, taking the following mass spectrum for the decay channel:
$m_{\eta^+}=300~\mathrm{GeV}$, $m_2=290~\mathrm{GeV}$, 
$m_H=220~\mathrm{GeV}$, $m_L=120~\mathrm{GeV}$,
$m_\psi=60~\mathrm{GeV}$, 
the constraints we have taken into account can be satisfied by choosing
suitable Yukawa couplings, $y^\nu,y^L$ and $y^R$.
Detailed signal and background analysis is necessary to estimate detectability of 
the cuspy feature. However, this is beyond the scope of this paper.


\subsection{Complex Scalar Dark Matter}
\begin{figure}[t]
\begin{center}
\includegraphics[scale=0.7]{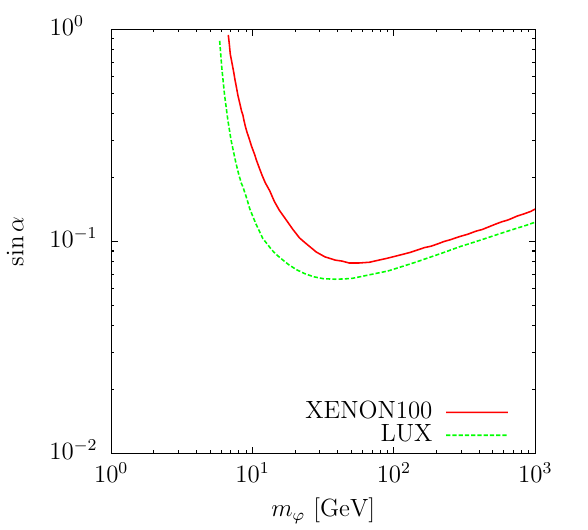}
\hspace{0.2cm}
\includegraphics[scale=0.7]{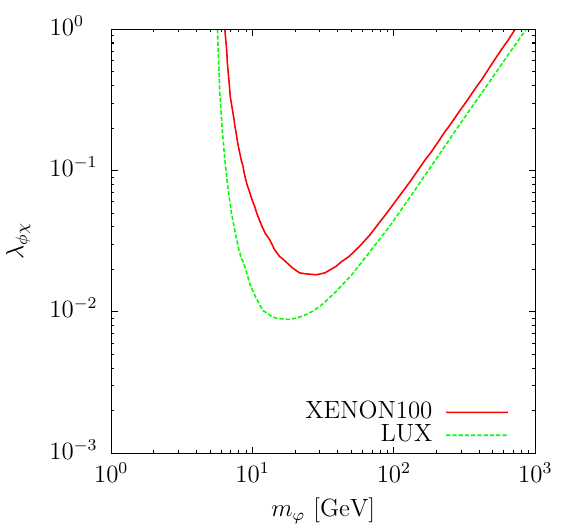}
\caption{Excluded parameter space in $m_\varphi-\sin\alpha$ plane
 (left panel) and $m_\varphi-\lambda_{\phi\chi}$ (right panel) by
 XENON100~\cite{Aprile:2012nq} and LUX~\cite{Akerib:2013tjd}. 
 The region above each line is excluded.}
\label{fig:dd_diagram}
\end{center}
\end{figure}

In this section, we discuss the case of complex scalar DM. 
A similar kind of $\mathbb{Z}_3$ scalar DM has been studied in
ref.~\cite{Belanger:2012vp, Belanger:2012zr}. 
First of all, we should take into account the severe constraint from
direct detection of DM. 
The complex scalar DM $\varphi\equiv\varphi_L$ with mass
$m_\varphi\equiv m_L$ has an interaction with the $Z$ boson,
since $\varphi$ includes a component of $SU(2)_L$ doublet $\eta^0$. 
Thus, much of the parameter space of the mixing angle is 
excluded by the direct detection experiments like
LUX~\cite{Akerib:2013tjd} and XENON100~\cite{Aprile:2012nq}. 
Here we investigate how small the mixing should be
 so as to evade the bound. The contribution to the spin
independent elastic cross section with nucleus comes from the $Z$ boson
exchange diagram, which is calculated as 
\begin{equation}
\sigma_{\mathrm{SI}}^Z=\frac{G_F^2}{2\pi}
\frac{m_\varphi^2m_A^2\sin^4\alpha}{\left(m_\varphi+m_A\right)^2}
\left[(A-Z)-Z(1-4\sin^2\theta_{W})\right]^2,
\label{eq:si}
\end{equation}
where $A$ is the mass number of nucleus. 
The elastic cross section with a proton is most stringently constrained by the LUX 
experiment. The excluded parameter space in
$m_{\varphi}-\sin\alpha$ plane is shown in the left panel in Fig.~\ref{fig:dd_diagram}. 
The requirement is roughly
$\sin\alpha\lesssim0.05$, and we take $\sin\alpha=0.05$ in the following discussion. 
Note that in the ordinary inert doublet model, the constraint from the $Z$ boson
exchange would be weakened by the existence of the mass
splitting between CP even and odd inert scalars. 

In addition, the scalar coupling $\lambda_{\phi\chi}$ is also relevant
for direct detection via Higgs boson exchange. 
The elastic cross section is calculated as 
\begin{equation}
 \sigma_{\mathrm{SI}}^h=
\frac{\lambda_{\phi\chi}^2}{4\pi}\frac{m_N^4}{\left(m_\varphi+m_N\right)^2m_h^4}
\left[(A-Z)C_n+ZC_p\right]^2,
\end{equation}
where the coefficients $C_p\approx C_n\approx0.29$ are calculated from
ref.~\cite{Belanger:2008sj}. 
The constraint on the coupling $\lambda_{\phi\chi}$ is shown in the
right panel in Fig.~\ref{fig:dd_diagram}. 
From this figure, we see that the coupling strength should be
$\lambda_{\phi\chi}\lesssim0.007$ so as not to conflict with the LUX result in
all the DM mass range. 
Moreover, DM with mass less than $100~\mathrm{GeV}$ tends to be excluded by the vacuum
conditions of Eqs.~(\ref{eq:vc1})$-$(\ref{eq:vc4}).
Note that the two contributions via $Z$ boson and Higgs boson
exchange should be combined together to do a thorough analysis, the
above discussion is sufficient to set a conservative limit.

\begin{figure}[t]
\begin{center}
\includegraphics[scale=0.8]{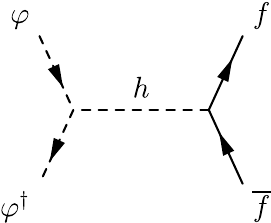}
\hspace{0.2cm}
\includegraphics[scale=0.8]{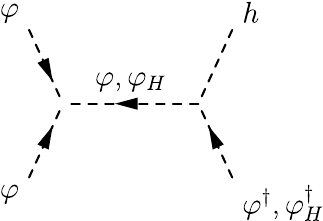}\\
\includegraphics[scale=0.8]{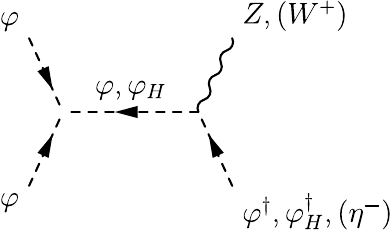}
\hspace{0.2cm}
\includegraphics[scale=0.8]{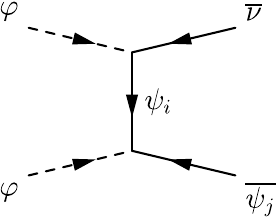}
\caption{Examples of (semi-)annihilation processes of complex scalar DM
 $\varphi$. For the s-channels written here, the corresponding
 t-channels also exist.}
\label{fig:2-body2}
\end{center}
\end{figure}

\begin{figure}[t]
\begin{center}
\includegraphics[scale=0.6]{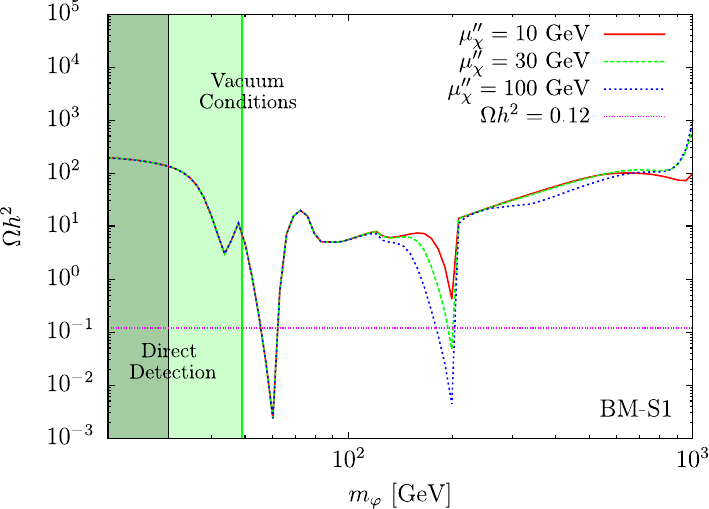}
\includegraphics[scale=0.6]{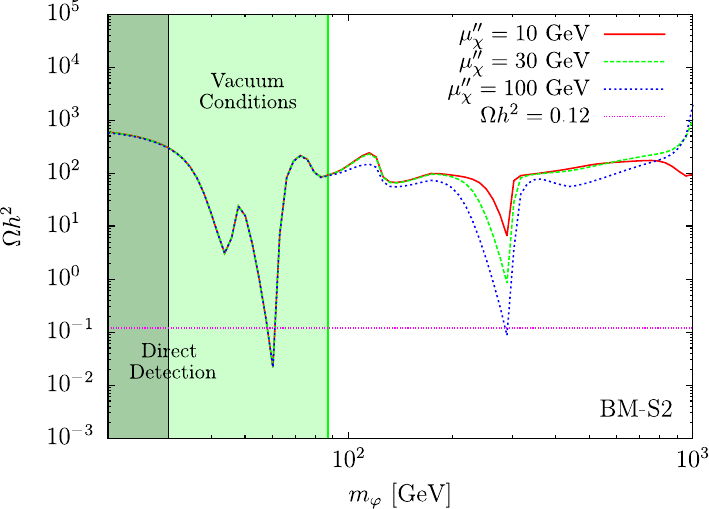}\\
\includegraphics[scale=0.6]{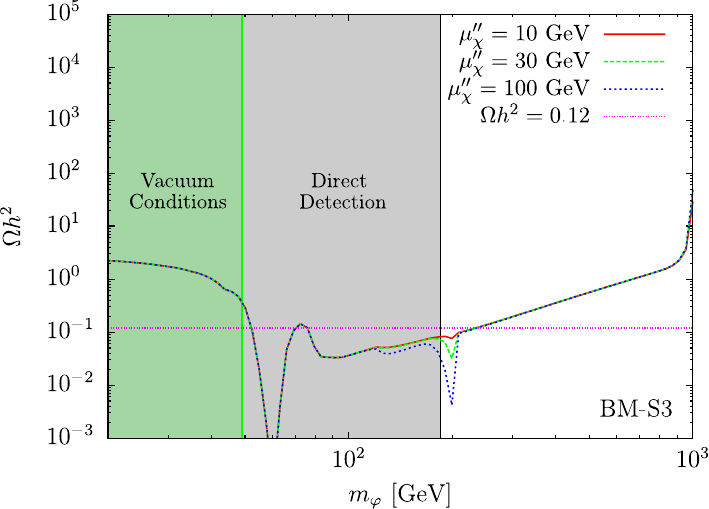}
\includegraphics[scale=0.6]{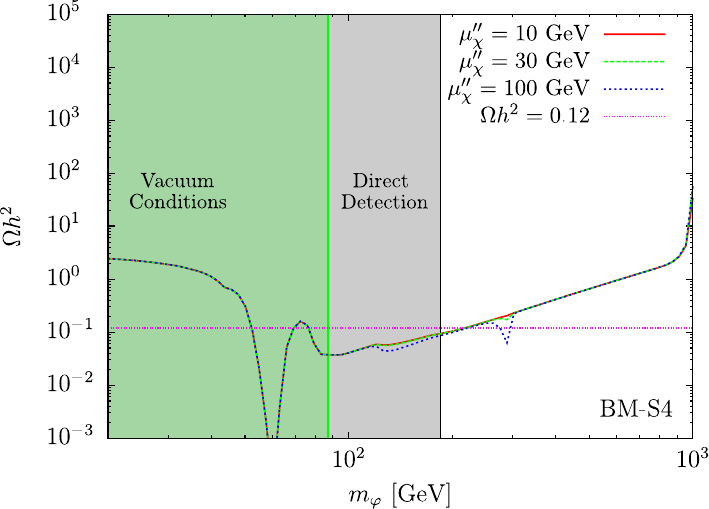}
\caption{DM mass dependence of relic density where the parameters are
 taken as in Tab.~\ref{tab:bm2}. Note that when one changes $\mu_\chi''$
 as in the figures, the parameters $y^L$ and $y^R$ also should be
 varied from the scale of the neutrino masses.} 
\label{fig:omega_b}
\end{center}
\end{figure}

Regarding annihilations of the scalar DM, there are many standard annihilation and
semi-annihilation channels such as $\varphi\varphi^{\dag}\to 
f\overline{f}$, $hh$, $ZZ$, $W^+W^-$ and $\varphi\varphi\to
h\varphi^\dag$, $Z\varphi^\dag$, $W^+\eta^-$ shown in
Fig.~\ref{fig:2-body2}. 
Since the mixing angle is strictly constrained in our case, the main
component of the scalar DM is $\chi$, rather than $\eta^0$. 
In Fig.~\ref{fig:omega_b} some plots of the DM mass dependence of the
relic density are shown for the parameter values given in Tab.~\ref{tab:bm2}. 
The lower bounds of the DM mass are obtained from the vacuum
condition Eq.~(\ref{eq:vc4}) as $m_\varphi\gtrsim49~\mathrm{GeV}$ for the left panels and
$m_\varphi\gtrsim87~\mathrm{GeV}$ for the right panels. 
In the upper panels, the region of light DM mass $12~\mathrm{GeV}\lesssim
m_\varphi\lesssim30~\mathrm{GeV}$ is ruled out by the direct
detection experiments, while the excluded mass range is $6~\mathrm{GeV}\lesssim
m_\varphi\lesssim185~\mathrm{GeV}$ in the lower panels. 
In all figures, the relic density of DM is drastically reduced around
$m_\varphi=m_h/2\approx63~\mathrm{GeV}$ due to the annihilation
channels $\varphi\varphi^\dag\to h\to b\overline{b}$, $W^*W$. 
Although the most of the (semi-)annihilations are suppressed for BM-S1
and BM-S2 (upper panels) due to the small $\lambda_{\phi\chi}$, the
damping around $200~\mathrm{GeV}$ in BM-S1 and 
$300~\mathrm{GeV}$ in BM-S2 come from the
semi-annihilation channels $\varphi\varphi\to \varphi_H\to
Z\varphi^\dag$, $h\varphi^\dag$. 
These channels are absent in the minimal singlet $\mathbb{Z}_3$ DM
model~\cite{Belanger:2012zr} since there is no second neutral $\mathbb{Z}_3$ scalar boson. 
These semi-annihilation channels have a significant contribution when the
cubic scalar coupling $\mu_\chi''$ is large. 
Another notable point in BM-S1 and BM-S2
is that the relic density of DM is a reduced
around $m_\varphi\approx1~\mathrm{TeV}$ 
for smaller cubic coupling $\mu_\chi''$
because the co-annihilation $\varphi\psi_i\to\overline{\psi_i}\to\eta^-\overline{\ell}$ 
is more effective. 

\begin{table}[t]
\begin{center}
\begin{tabular}{|c||c|c|c|c|c|c|c|c|}\hline
 & $m_H-m_\varphi~[\mathrm{GeV}]$ &
 $\lambda_{\phi\chi}$ & $\lambda_{\eta\chi}$ & $\lambda_2$ & $\lambda_3$ & 
 $\lambda_4$\\
\hhline{|=#=|=|=|=|=|=|}
 BM-S1 & $200$ & $0.01$ & $1.0$ & $0.1$ & $0.5$ & $0.5$\\\hline
 BM-S2 & $300$ & $0.01$ & $1.0$ & $0.1$ & $0.5$ & $0.5$\\\hline
 BM-S3 & $200$ & $0.1$  & $1.0$ & $0.1$ & $0.5$ & $0.5$\\\hline
 BM-S4 & $300$ & $0.1$  & $1.0$ & $0.1$ & $0.5$ & $0.5$\\\hline
\end{tabular}
\caption{Parameter sets of benchmark point for the complex scalar
 DM. The other parameters are set to $\sin\alpha=0.05$, 
 $y_{i\alpha}^\nu=0.01$, $m_i=1~\mathrm{TeV}$ and $\mu_\chi''(y_{ij}^L+y_{ij}^R)=10~\mathrm{GeV}$.}
\label{tab:bm2}
\end{center}
\end{table}

For BM-S3 and BM-S4 (lower panels), while the DM mass is strongly
constrained by direct detection, the contributions 
of (semi-)annihilations through 
the coupling $\lambda_{\phi\chi}$ can be large and the relic density
is much reduced. Thus the $\mu_\chi''$ dependence becomes relatively
smaller than the BM-S1 and BM-S2 cases. Moreover 
the DM mass scale of several hundred GeV can be detected by 
future direct detection experiments like XENON1T. 

We have performed an investigation of the case with large Yukawa coupling $y^\nu$
 which is similar to the Dirac DM case.
 In this case, when $y^L,y^R\sim\mathcal{O}(1)$ the smaller
$\mu_\chi''$ is required to derive the 
known neutrino mass scale. As a result, the damping around 
$m_\varphi\approx 200~\mathrm{GeV}$
in BM-S1 and BM-S3 and $300~\mathrm{GeV}$ in BM-S2 and
BM-S4 almost disappear. Instead of that, the semi-annihilation channel
$\varphi\varphi\to\overline{\nu}\overline{\psi_i}$ depicted in the lower
right diagram of Fig.~\ref{fig:2-body2} affects the DM relic density
when the DM is heavier than $m_\psi/2$. 

As already discussed, although much of the parameter space has already
been ruled out by the direct searches for DM, 
there remains the possibility to detect the complex scalar DM by XENON1T.
However, there are essentially no differences from typical $\mathbb{Z}_2$ symmetric DM. 
For indirect searches, while the scalar DM basically produces some SM
particles such as $W$, $Z$, $h$, $\ell$, $\nu$, 
it would be difficult to see any characteristic signatures. 
The annihilation channel $\varphi^\dag\varphi\to\nu\overline{\nu}$ via
$\psi_i$ exchange is suppressed by the small mixing angle $\sin\alpha$ and
it is p-wave suppressed, which is different from the Dirac
DM case. Thus even if we assume a large Yukawa coupling $y^\nu$, the
annihilation cross section is too small to observe, and we have only one
monochromatic neutrino emission from the semi-annihilation
$\varphi\varphi\to\overline{\nu}\overline{\psi_i}$ in the right bottom
process in Fig.~\ref{fig:2-body2}.
However, two monochromatic neutrinos would be seen at
$E_\nu=m_\varphi\left(1-m_i^2/(4m_\varphi^2)\right)$ 
from semi-annihilation if the masses of
$\psi_1$ and $\psi_2$ are different. 
This process does not exist in
the similar $\mathbb{Z}_3$ scalar DM model of ref.~\cite{Belanger:2012vp}.

\section{Summary and Conclusions}
\label{sec:4}
We have considered a model with $\mathbb{Z}_3$ symmetry. 
In this model, the neutrino masses are generated at
the two-loop level and the known neutrino mass scale has been derived with 
a reasonable value for the coupling strength. 
The DM in the model is either a Dirac fermion or a complex scalar as a result of 
the exact $\mathbb{Z}_3$ symmetry. 
The semi-annihilation processes are important to reduce the relic density
effectively in the early universe for both of DM particles. 
We have discussed the DM relic density in the two
cases of the small and large Yukawa couplings because of the LFV constraint. 
In particular for the Dirac DM $\psi$ with the small Yukawa coupling, 
although the standard annihilation channel is suppressed, the DM relic density
can be compatible with the observed value due to the semi-annihilation processes. 
Direct detection constrains the complex scalar DM $\varphi$ to be dominantly singlet.
The semi-annihilation processes for the scalar DM are controlled
by the cubic coupling $\mu_\chi''$ and influence the DM relic density. 

The $\mathbb{Z}_3$ symmetric Dirac DM in this model potentially has some interesting signatures, which may be detected by indirect detection and colliders.
In particular, for the case with large Yukawa coupling, 
two monochromatic neutrinos may be observed from the annihilation 
$\psi\overline{\psi}\to\nu\overline{\nu}$ and the semi-annihilation
$\psi\psi\to\nu\overline{\psi}$ since these cross sections can be the same order
of magnitude.  The double peak of neutrino flux may be detected by
neutrino observatories such as IceCube. 
In addition, in the decay of the $\mathbb{Z}_3$ charged boson $\eta^+$, 
the cusp feature of the invariant mass distribution 
may be seen at collider experiments. 
For the complex scalar DM, if the masses of $\psi_1$ and $\psi_2$ are
different, two monochromatic neutrinos would be emitted from the semi-annihilation
$\varphi\varphi\to\overline{\nu}\overline{\psi_i}$. 


\section*{Acknowledgments}
The authors would like to thank Christopher McCabe for careful reading
of the manuscript.
The work of M.~A. is supported in part by the Grant-in-Aid for
Scientific Research, Nos. 25400250 and 26105509. 
T.~T. acknowledges support from the European ITN
project (FP7-PEOPLE-2011-ITN,
PITN-GA-2011-289442-INVISIBLES). T.~T. thanks Kanazawa University for
the travel support and local hospitality during some parts of this work. 



\end{document}